\documentclass[sigconf]{acmart}
\acmSubmissionID{906}

\usepackage{booktabs} % For formal tables

\usepackage{bm}
\usepackage{wrapfig}
\usepackage{graphicx}
\usepackage[noabbrev, capitalise]{cleveref}
\usepackage{xcolor,colortbl}
\newcommand{\E}{\mathop{\mathbb{E}}}

\renewcommand{\d}{\mathrm{d}}
\newcommand{\rv}[1]{\mathbf{#1}}
\newcommand{\rd}[1]{\mathcal{#1}}

\newcommand{\var}{\text{Var}}
\newcommand{\<}{\langle}
\renewcommand{\>}{\rangle}
\newcommand{\D}{\Delta}
\newcommand{\p}{\partial}

\newcommand{\diffest}{\< \D F_i \>}
\newcommand{\est}{\< F_i \>}
\newcommand{\varest}{\var [\< F_i \>]}
\newcommand{\vardiff}{\var [\< \D F_i \>]}
\newcommand{\varmeta}{\var [\< F_i \>_M]}
\newcommand{\varprev}{\var [\< F_{i-1} \>_M]}

\newcommand{\newtext}[1]{#1}
\newcommand{\ie}{i.e.,\ }
\newcommand{\eg}{e.g.,\ }

\usepackage[ruled]{algorithm2e} % For algorithms

\SetAlFnt{\small}
\SetAlCapFnt{\small}
\SetAlCapNameFnt{\small}
\SetAlCapHSkip{0pt}

\SetKwComment{Comment}{// }{}
\newcommand{\cnote}[2]{\textit{#1} \copyright{} 2023 #2}

\acmJournal{TOG}
\copyrightyear{2023}
\acmYear{2023}
\setcopyright{rightsretained}
\acmConference[SA Conference Papers '23]{SIGGRAPH Asia 2023 Conference Papers}{December 12--15, 2023}{Sydney, NSW, Australia}
\acmBooktitle{SIGGRAPH Asia 2023 Conference Papers (SA Conference Papers '23), December 12--15, 2023, Sydney, NSW, Australia}
\acmDOI{10.1145/3610548.3618244}
\acmISBN{979-8-4007-0315-7/23/12}

\begin{document}
% Title portion
\title{Joint Sampling and Optimisation for Inverse Rendering}

\author{Martin B\'alint}
\orcid{0000-0001-6689-4770}
\affiliation{%
 \institution{Max Planck Institute for Informatics}
 \streetaddress{Campus E1 4}
 \city{Saarbruecken}
 \postcode{66123}
 \country{Germany}}
\email{mbalint@mpi-inf.mpg.de}

\author{Karol Myszkowski}
\orcid{0000-0002-8505-4141}
\affiliation{%
 \institution{Max Planck Institute for Informatics}
 \streetaddress{Campus E1 4}
 \city{Saarbruecken}
 \postcode{66123}
 \country{Germany}}
\email{karol@mpi-inf.mpg.de}

\author{Hans-Peter Seidel}
\orcid{0000-0002-1343-8613}
\affiliation{%
 \institution{Max Planck Institute for Informatics}
 \streetaddress{Campus E1 4}
 \city{Saarbruecken}
 \postcode{66123}
 \country{Germany}}
\email{hpseidel@mpi-inf.mpg.de}

\author{Gurprit Singh}
\orcid{0000-0003-0970-5835}
\affiliation{%
 \institution{Max Planck Institute for Informatics}
 \streetaddress{Campus E1 4}
 \city{Saarbruecken}
 \postcode{66123}
 \country{Germany}}
\email{gsingh@mpi-inf.mpg.de}

\begin{teaserfigure}
  \centering
  \includegraphics[width=\textwidth]{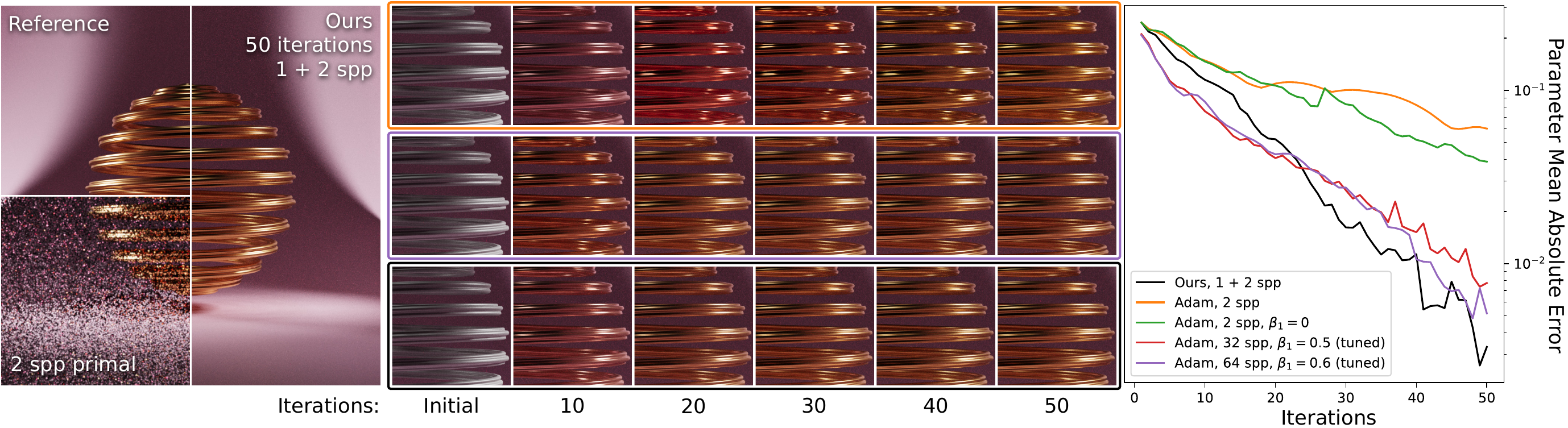}
  \caption{We simultaneously optimise the spiral's base colour, metalness and roughness. Our gradient \textit{meta-estimator} quickly recovers the gradient with just one \newtext{finite-difference} and two proportional samples per pixel. Adam's default first moment estimator ($\beta_1 = 0.9$) cannot adapt as quickly and often overshoots. After tuning Adam's hyperparameters for this specific problem, it approaches out method at 32 spp and matches it only at 64 spp. \cnote{Banner 2}{Wenzel Jakob}}
  \label{fig:exp2}
\end{teaserfigure}

\begin{abstract}
When dealing with difficult inverse problems such as \newtext{inverse rendering}, using Monte Carlo \newtext{estimated gradients} to optimise parameters \newtext{can slow down convergence due to variance.} Averaging many \newtext{gradient} samples in each iteration reduces this variance trivially. However, for problems that require thousands of optimisation iterations, the computational cost \newtext{of this approach rises quickly.}

We derive a theoretical framework for interleaving sampling and optimisation. We \newtext{update and} reuse past samples \newtext{with low-variance finite-difference} estimators that describe the change in the estimated gradients between each iteration. By combining \newtext{proportional} and \newtext{finite-difference} samples, we continuously reduce the variance of our novel gradient meta-estimators throughout the optimisation process. We investigate how our estimator interlinks with Adam and derive a stable combination.

We implement our method for \newtext{inverse path tracing} and demonstrate how our estimator speeds up convergence \newtext{on difficult} optimisation tasks.
\end{abstract}



\maketitle

\section{Introduction}

Forward rendering involves solving the light transport integrals with given scene parameters,~\eg geometry, materials, textures, by numerically estimating the rendering equation \cite{kajiya1986rendering,PBRT3e}.
Inverse rendering reverses this \newtext{process by estimating the} scene parameters starting from a given target image. This process involves inverting the rendering equation~\cite{kajiya1986rendering}.

Such inversion tasks are typically solved by gradient descent. Physically-based differentiable renderers \cite{Mitsuba3,li2018differentiable,zhang2020psdr} facilitate these gradient-based optimisation methods.
\newtext{The process involves backpropagating from an underlying loss function, quantifying the disparity between an image generated with the current parameters and the target image, resulting in gradients w.r.t. the scene parameters. These gradient values are approximated from a given set of samples, and subsequently, the scene parameters are adjusted using these gradients to minimise the loss. The ultimate goal is to converge to a parameter set that produces the target image.}
Due to the nature of Monte Carlo integration, the estimated gradients can be extremely noisy, hampering the performance of gradient-based optimisers.
In inverse rendering, gradients are estimated with tens to hundreds of rays traced per pixel \cite{NimierDavid2020Radiative, zhang2019differential} to minimise noise.
Usually, inverse rendering requires hundreds to thousands of iterations to converge; recomputing these gradient estimates in every iteration \newtext{comes at a large cost.}

In this paper, we propose a theoretical framework that jointly considers sampling and optimisation by deriving a theoretical framework for interleaving them. 
We reuse past samples without introducing bias thanks to finite-\newtext{difference} estimators that describe the change in the estimated gradients between each iteration. 
By combining proportional and \newtext{finite-difference} samples, we continuously reduce the variance of our novel gradient
meta-estimators throughout the optimisation process. 

First, we introduce our meta-estimation theory and then discuss our variance estimation strategies used to derive coefficients for our meta-estimator. We investigate how our estimator interlinks with Adam and derive a stable combination. We run experiments to evaluate our method in the context of inverse rendering. Finally, we discuss our method concerning future and concurrent work and give our conclusions.

Our contributions include:
\begin{itemize}
    \item Meta-estimation theory on combining proportional and finite-\newtext{difference} estimators.
    \item Practical variance \newtext{approximation} techniques to effectively implement meta-estimation.
    \item Implementation and evaluation of meta-estimation for inverse rendering. (We will release our code upon acceptance.)
\end{itemize}
%

%%%%%%%%%%%%%%%%%%%%%%%%%%%%%%%%%%%%%%
\section{Related work}
%%%%%%%%%%%%%%%%%%%%%%%%%%%%%%%%%%%%%%

%%%%%%%%%%%%%%%%%%%%%%%%%%%%%%%%%%%%%%
\paragraph{Differentiable Path Tracing}
%%%%%%%%%%%%%%%%%%%%%%%%%%%%%%%%%%%%%%

Path tracing accounts for global illumination through physically accurate light transport by Monte Carlo integration of the rendering equation \cite{kajiya1986rendering}. Recent works proposed various approaches to differentiate such Monte Carlo integrals and estimate derivatives w.r.t. scene parameters. \cite{NimierDavid2020Radiative, Zeltner2021MonteCarlo, Mitsuba3,li2018differentiable,zhang2020psdr}. While our work applies to any method using gradient descent on Monte Carlo estimated gradients, we mainly experiment with Path Replay Backpropagation \cite{Vicini2021PathReplay}; a well-established state-of-the-art method for inverse path tracing, implemented in Mitsuba 3 \cite{Mitsuba3}.

\newtext{Previous works have focused on sampling strategies \cite{zhang2021antithetic, yan2022efficient, Bangaru2020Unbiased} and improving the optimisation itself \cite{NimierDavid2020Radiative,Vicini2021PathReplay} to reduce noise in the gradients. Particularly relevant, concurrent work by \citet{chang2023parameter} applies ReSTIR \cite{bitterli20spatiotemporal} in parameter-space with the same goal of reducing the variance of the estimated gradients.}

%%%%%%%%%%%%%%%%%%%%%%%%%%%%%%%%%%%%%%
\paragraph{Ray Differentials}
%%%%%%%%%%%%%%%%%%%%%%%%%%%%%%%%%%%%%%
\citet{igehy1999tracing} first proposed ray differentials to approximate derivatives for texture interpolation and anisotropic filtering. \citet{kettunen2015gradient} combine ray differentials with gradient-domain MLT \cite{lehtinen2013gradient} to build unbiased image-space gradient estimators for gradient-domain path tracing. \citet{manzi2016temporal} extend their work to the spatiotemporal domain.

We apply the general idea of finite-\newtext{difference} estimation to temporal gradient averaging on a set of parameters. As we do not assume any structure between individual parameters, we forgo Poisson reconstruction and instead statistically average proportional and integrate \newtext{finite-difference} samples.

%%%%%%%%%%%%%%%%%%%%%%%%%%%%%%%%%%%%%%
\paragraph{Gradient averaging} 
%%%%%%%%%%%%%%%%%%%%%%%%%%%%%%%%%%%%%%
Iterating with the arithmetic mean of gradient samples is well-understood to improve the convergence of optimisers. Several recursive schemes \cite{nesterov1983method, polyak1992acceleration} achieve fast convergence on convex problems \cite{Moulines2011non}, with some proving particularly useful in deep learning \cite{sutskever2013importance}. \citet{kingma2014adam} propose start-up bias-corrected exponential moving averaging on gradients; Adam remains the de-facto optimisation algorithm for deep learning and inverse rendering applications.

In recent work, \citet{gower2020variance} analyse gradient averaging methods based on finite sums; they show improvements in convergence analogous to our work, although limited to convex problems. Unfortunately, the finite-sum setting of algorithms like SAGA \cite{defazio2014adv} and SVRG \cite{johnson2013acc} does not generalise to Monte Carlo integration \cite{Nicolet2023Recursive}.

\newtext{
Reducing the gradient variance is well understood to improve convergence speed and stability. Previous works on optimising neural networks increase the batch size to reduce this variance, which is often preferable over slower learning rates \cite{smith2018dont}.
}

%%%%%%%%%%%%%%%%%%%%%%%%%%%%%%%%%%%%%%
\paragraph{Control Variates} 
%%%%%%%%%%%%%%%%%%%%%%%%%%%%%%%%%%%%%%
\citet{fieller1954sampling} first propose control variates as a weighted combination of correlated estimators, one of which must be of a closed-form integral. \citet{owen2013monte} shows that the optimal control weight is proportional to the covariance of the estimators. \citet{rousselle2016image} generalise control variates to any pair of correlated estimators through two-level Monte Carlo integration; they apply their work to spatiotemporal gradient-domain rendering. Concurrent with our work, \citet{Nicolet2023Recursive} further generalise control variates to recursive estimation, applying it to primal renderings in the context of inverse path tracing.

Our work is distinctively different from control variates in that we build on an independent finite-\newtext{difference} estimator rather than a pair of correlated estimators. In particular, this formulation lets us avoid covariance terms in our weighting scheme. 

%%%%%%%%%%%%%%%%%%%%%%%%%%%%%%%%%%%%%%
\section{Differential Meta-Estimators}
%%%%%%%%%%%%%%%%%%%%%%%%%%%%%%%%%%%%%%
Various Monte Carlo methods estimate a sequence of integrals. Often, each integral is a function of the previous one, with the sequence converging to a solution. Optimisation via inverse Monte Carlo is a prime example; we estimate gradients in each iteration, adjust parameters accordingly, and repeat the process. 

\newtext{
Our work is focused on improving the convergence speed and stability of the optimisation process by reducing the variance of the estimated gradients. We draw inspiration from control theory, specifically noise reduction through the combination of proportional and differential signals. These methods assume that samples are drawn from known probability distributions, usually normal distributions with known variances~\cite{Kalman}. Unfortunately, we cannot make such assumptions when dealing with Monte Carlo noise.
}

\newtext{
We combine two estimators: a \textit{proportional estimator} $\est$ --- any Monte Carlo gradient estimator sampled independently between iterations --- and a \textit{finite-difference estimator} $\diffest$ that estimates the change of a gradient between two consecutive iterations.
}

%%%%%%%%%%%%%%%%%%%%%%%%%%%%%%%%%%%%%%
\paragraph{Notation} 
%%%%%%%%%%%%%%%%%%%%%%%%%%%%%%%%%%%%%%
Let $F_i$ denote the integral of function $f$ over the domain $\rd{X}$, given parameters $\pi_i$ for the current iteration $i \in [0, \infty)$:
\begin{equation}
    \label{eq:F_def}
    F_i = \int_{\rd{X}} f(\rv{x}, \pi_i) \d \rv{x} \text{ .}
\end{equation}
Let $\est$ denotes the (proportional) Monte Carlo estimator of $F_i$, meaning $\E[\est] = F_i$. 
For example, an estimator may sample $f$ given a density $p$ over $\rd{X}$:
\begin{equation}
    \label{eq:est_def}
    \est = \frac{f(\rv{x}, \pi_i)}{p(\rv{x}, \pi_i)} \text{ .}
\end{equation}

\begin{figure*}[ht!]
  \centering
  \includegraphics[width=\textwidth]{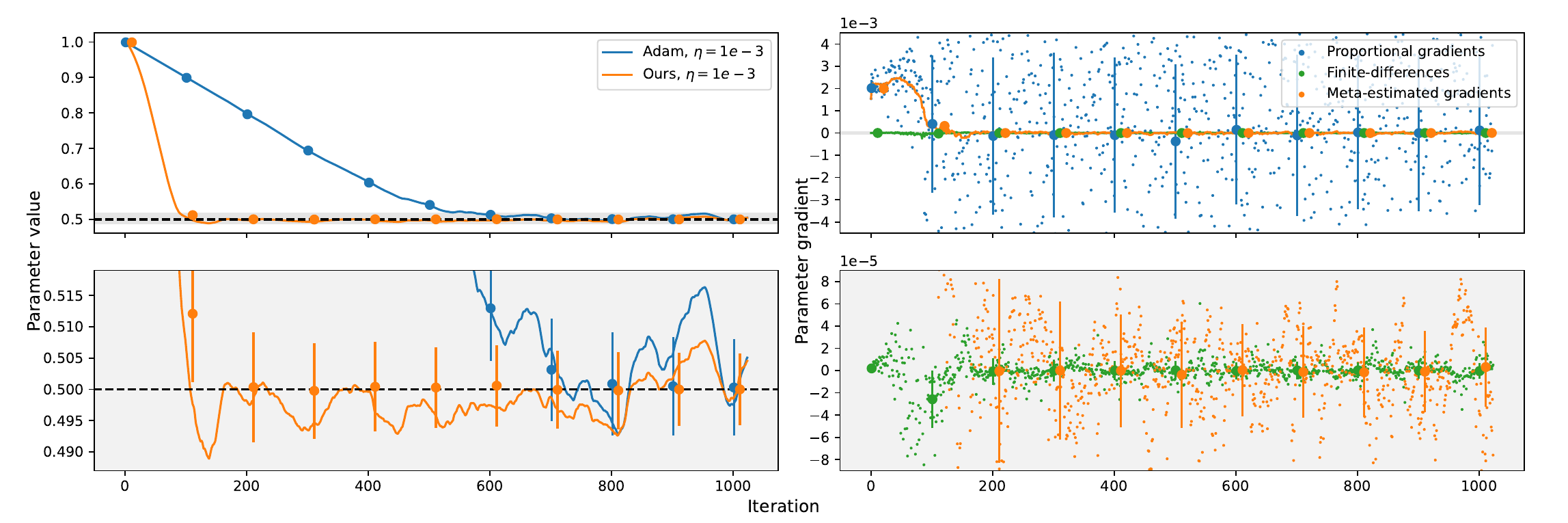}
  \caption{\newtext{
  We optimise the rate parameter of an exponential distribution such that the mean of the distribution matches our target value of 2.0. We take 32 samples of the distribution in each iteration and compute an L2 loss between their mean and the target value. 
  The bottom row shows insets of the graphs in the top row, indicated by grey regions. 
  On the left, we show how Adam and our method reach the ground truth rate parameter 0.5. 
  Error bars show the run-to-run variation of the optimised parameter. 
  Our method converges significantly faster and is more stable than Adam. 
  On the right, we show the estimators we use for our method; the proportional estimator has a large variance, while the finite-difference estimator is much less noisy. Our meta-estimator combines both, with its variance reducing over time.
  }}
  \label{fig:exposition}
\end{figure*}

\newtext{
\paragraph{Finite-difference estimation} We write the change of $F_i$ between consecutive steps as:
}
\begin{equation}
    \label{eq:diff_def}
    \D F_i = F_i - F_{i-1} \text{ .}
\end{equation}
\newtext{
A finite-difference estimator ($\diffest$) estimates this change, ideally with a low variance. For example, we can substitute \cref{eq:F_def} into \cref{eq:diff_def}:
}
\begin{equation}
    \D F_i = \int_{\rd{X}} f(\rv{x}, \pi_i) - f(\rv{x}, \pi_{i-1}) \d \rv{x} \text{ ,}
\end{equation}
\newtext{
and sample with a density $p$ like in \cref{eq:est_def}:
}
\begin{equation}
    \label{eq:est_triv}
    \diffest = \frac{f(\rv{x}, \pi_i) - f(\rv{x}, \pi_{i-1})}{p(\rv{x}, \pi_i)} \text{ .}
\end{equation}
\newtext{
Here we assume that $f$ is continuous w.r.t. $\pi_i$.
Although our theory may apply to any Monte Carlo integral $F_i$, we analyse the case where $f(x, \pi_i) = \partial L / \partial \pi_i$ is the gradient at the $i$-th iteration, for some objective $L$. 
}

%%%%%%%%%%%%%%%%%%%%%%%%%%%%%%%%%%%%%%
\subsection{Meta-estimation} 
\label{subsec:meta_estimation}
%%%%%%%%%%%%%%%%%%%%%%%%%%%%%%%%%%%%%%

\newtext{Our \textit{meta-estimator} aims to optimally combine each proportional $\est$ and finite-difference  estimator $\diffest$ available until the current step $i$. 
In this subsection, we establish the theoretical conditions required for a variance-optimal, unbiased combination of both estimators.
}

\newtext{
A finite-difference estimator $\diffest$, by its definition in \cref{eq:diff_def}, lets us update any proportional estimator from the previous step ($\< F_{i-1} \>$)  to the current step $i$. This update can be done simply by addition without introducing any bias. We can easily show this by expanding the expected value of the sum:
}
\newtext{\begin{multline}
    \E[\< F_{i-1} \> + \diffest] = \\
    \E[\< F_{i-1} \>] + \E[\diffest] = F_{i-1} + F_i - F_{i-1} = F_i \text{ .}
\end{multline}}
\newtext{
We define our meta estimator as a weighted sum over the combination of all previous estimators up until step $i$-1~\ie $\< F_{i-1} \>_M$ and the current proportional and finite-difference estimators: 
}
\begin{equation}
    \label{eq:recurrent_meta}
    \boxed{\est_M = \alpha_i ( \< F_{i-1} \>_M + \diffest) + (1 - \alpha_i) \est \text{ .}}
\end{equation}
We initialise $\< F_0 \>_M = \< F_0 \>$. As we sample all $\est$ and $\diffest$ independently, the optimal $\alpha_i$ coefficients are given by inverse variance weighting~\cite{sinha2011statistical}:
\begin{equation}
    \label{eq:alpha}
    \alpha_i = \frac{\varest}{\varest + \varprev + \vardiff} \text{ .}
\end{equation}
This simple recurrent relation captures the variance optimal combination of all previously sampled proportional and finite-difference estimators.

%%%%%%%%%%%%%%%%%%%%%%%%%%%%%%%%%%%%%%
%\paragraph{Discussion}
%%%%%%%%%%%%%%%%%%%%%%%%%%%%%%%%%%%%%%
\newtext{
To summarise, we introduce the optimal and unbiased meta-estimator in \cref{eq:recurrent_meta}. 
However, in practice, we use a more efficient implementation which suffers from some start-up bias. We describe this version in the following section. To visualise the estimators mentioned above and to motivate the design of our optimiser, we show a simple example in \cref{fig:exposition}.
}

%%%%%%%%%%%%%%%%%%%%%%%%%%%%
\section{Variance estimation}
%%%%%%%%%%%%%%%%%%%%%%%%%%%%
\newtext{
Implementing~\cref{eq:recurrent_meta} in practice presents a challenge due to the unknown variances of our estimators. In this section, we describe how we approximate each variance term. We must balance three main objectives: the efficiency of our variance approximation methods, the optimality of our approximated $\alpha_i$ coefficients, and any bias potentially introduced to $\est_M$.
}

%%%%%%%%%%%%%%%%%%%%%%%%%%%%
\subsection{\newtext{Proportional estimator variance}}
%%%%%%%%%%%%%%%%%%%%%%%%%%%%

\newtext{We approximate $\varest$ as a zero-centred raw moment \cite{papoulis1984probability}, computed using an \newtext{\textit{exponential moving average} (EMA)} with coefficient $\beta_F$}:
\begin{equation}
    \label{eq:varest}
    \varest = \beta_F \, \var [\< F_{i-1} \>] + (1 - \beta_F) \est^2 \text{ .}
\end{equation}
\newtext{This formulation is similar to Adam's second moment estimate \cite{kingma2014adam}.
Here, $\varest$ is a large, stable value that only varies in the initial stage of optimisation, when parameter changes can notably affect the problem's overall noise characteristics. 
A large $\beta_F$ coefficient minimises the correlation of the approximate variance to any singular $\est$, resulting in an overall stable variance approximation.}

\subsection{\newtext{Finite-difference estimator variance}}
\newtext{
As the proportional estimator reaches steady-state, we can safely assume that $\est$ are identically distributed over consecutive iterations. 
Unfortunately, the same observation does not apply to finite-difference estimation as this finite-difference depends on the optimisation step we take in the previous iteration. 
For example, a larger step will cause a larger shift in the per-parameter gradients.
}

\newtext{
To resolve this issue we propose to decouple the optimisation step size ($|\D \pi_i|$) from the approximated finite-difference estimator variance ($\vardiff$).
We begin the derivation of this decoupling by expanding the fraction in \cref{eq:est_triv} by the Euclidean step size $||\D \pi_i||_2$ of the previous iteration:
}
\begin{multline}
    \label{eq:vardiff_h1}
    \diffest = \frac{f(\rv{x}, \pi_i) - f(\rv{x}, \pi_{i-1})}{p(\rv{x}, \pi_i)} = \frac{f(\rv{x}, \pi_i) - f(\rv{x}, \pi_{i-1})}{||\D \pi_i||_2 p(\rv{x}, \pi_i)} ||\D \pi_i||_2 \,.
\end{multline}
\newtext{
For sufficiently small step sizes,  we can rearrange the terms in~\cref{eq:vardiff_h1} to approximate the finite-difference of gradients $f$ as the second-order gradient ($\partial f / \partial \pi$), times a unit-directional vector, times the left over terms: 
\begin{align}
\label{eq:vardiff_h}
\diffest  
    \approx \left( \frac{\p f}{\p \pi} (\rv{x}, \pi_i) \cdot \frac{\D \pi_i}{||\D \pi_i||_2} \right)  \frac{||\D \pi_i||_2}{p(\rv{x}, \pi_i)} \text{ .}
\end{align}
Applying the variance operator to~\cref{eq:vardiff_h} gives us the decoupled finite-difference variance ($\vardiff_D$):
}
\begin{multline}
    \vardiff \approx \var \left[  \left( \frac{\p f}{\p \pi} (\rv{x}, \pi_i) \cdot \frac{\D \pi_i}{||\D \pi_i||_2} \right)  \frac{1}{p(\rv{x}, \pi_i)} \right] ||\D \pi||_2^2  \\
    = \vardiff_D ||\D \pi||_2^2 \text{ .}
\end{multline}
We use a zero-centred EMA, with a coefficient $\beta_\D$, to approximate this decoupled variance as:
\begin{equation}
    \vardiff_D = \beta_\D \, \var [\< \D F_{i-1} \>]_D + (1 - \beta_\D) \frac{\diffest^2}{||\D \pi||_2^2} \text{ .}
\end{equation}
We generally use a small $\beta_\D$ coefficient as $\vardiff_D$ can change quickly and is typically less noisy than $\varest$. Finally, we can rescale $\vardiff_D$ to estimate $\vardiff$:
\begin{equation}
    \label{eq:vardiff}
    \vardiff = \vardiff_D \, ||\D \pi||_2^2 \text{ .}
\end{equation}
\newtext{We found this decoupled variance more closely distributed between iterations, better suited for approximation via moving averages.} 

%%%%%%%%%%%%%%%%%%%%%%%%%%%%%%%%%%%%%%%%%%%%%%%
\subsection{\newtext{Meta-estimator variance}}
%%%%%%%%%%%%%%%%%%%%%%%%%%%%%%%%%%%%%%%%%%%%%%%%
We \newtext{approximate} the variance of our meta estimator in~\cref{eq:recurrent_meta} by  recurrently applying the variance operator: 
\begin{multline}
    \label{eq:varmeta}
    \varmeta = \alpha_i^2 ( \varprev + \vardiff)  \\
    + (1 - \alpha_i)^2 \varest \text{ .}
\end{multline}
\newtext{Here we assume $\alpha_i$ to be a non-random value to simplify our mathematical derivation. Later in~\cref{sec:experiments}, we show the choice of $\alpha_i$ is less significant as long as its correlation with the gradient samples diminishes.}

%%%%%%%%%%%%%%%%%%%%%%%%%%%%
\subsection{Alpha clipping}
%%%%%%%%%%%%%%%%%%%%%%%%%%%%
Meta-estimation is most vulnerable to underestimated $\varmeta$; unless a significant $\vardiff$ indicates a shift, $\est_M$ will only slowly correct its overconfidently estimated value by averaging $\est$ over many iterations. 
The risk of underestimation is the greatest while our exponential moving averages accumulate their initial samples. Clipping alpha based on the iteration resolves this risk:
\begin{equation}
    \label{eq:alpha_clip}
    \alpha_i = \min (\alpha_i, 1 - 1 / (i + 1)) \text{ .}
\end{equation}
Intuitively,~\cref{eq:alpha_clip} constrains alpha by the perfect average of all previous estimates; any value above this must be overestimated. We generalise this observation to the entire optimisation process, assuming that $\varest$ is similar in subsequent steps:
\begin{equation}
    \label{eq:alpha_clip_2}
    \alpha_i = \min (\alpha_i, 1 / (2 - \alpha_{i-1}) ) \text{ .}
\end{equation}
%

%%%%%%%%%%%%%%%%%%%%%%%%%%%%
\section{Optimisation}
%%%%%%%%%%%%%%%%%%%%%%%%%%%%
\newtext{
Combining meta-estimation and optimisation creates a complex feedback loop; the meta-estimated gradients $\est_M$ depend on their finite-difference estimates $\diffest$, which depend on the optimiser's steps $\D \pi_i$, which, in turn, depend on the gradients estimated in the previous step $\< F_{i-1} \>_M$. It becomes crucial that the meta-estimator provides the optimiser with reliable gradients and that the optimiser makes steps that let the meta-estimator converge. We aim to combine Adam with meta-estimated gradients. We explain Adam's variance approximation and update rule to show where we can integrate meta-estimation.}

Adam \cite{kingma2014adam} is well known for its robustness to outlier gradient samples; upon encountering an outlier, the estimated second moments adjust in the same step, swiftly pulling down the step size. This mechanism works because Adam first updates its second-moment estimate:
\begin{equation}
    \label{eq:adam_moment}
    v_i = \beta_2 v_{i-1} + (1 - \beta_2) \est^2 \text{ ,}
\end{equation}
corrects the EMA startup bias:
\begin{equation}
    \label{eq:adam_ema_startup}
    \hat{v_i} = v_i / (1 - \beta_2^i) \text{ ,}
\end{equation}
and then divides the step size by its square root:
\begin{equation}
    \label{eq:adam_step}
    \D \pi_{i+1} = - \eta \frac{\hat{m_i}}{\sqrt{\hat{v_i}} + \epsilon} \text{ ,}
\end{equation}
where $m_i$ and $v_i$ refer to Adam's moment estimates, $\eta$ to the learning rate, $\beta_2$ to Adam's second moment coefficient, and $\epsilon$ to a small value to ensure numerical stability. $\alpha_i$ and $\varest$ behave similarly in our case; first, we update $\varest$ for the current step (\cref{eq:varest}), compute $\alpha_i$ (\cref{eq:alpha}), and add the outlier gradient $\est$ to our meta-estimator $\est_M$ weighted by $(1 - \alpha_i)$ (\cref{eq:recurrent_meta}). Therefore, just like $\beta_2$ for Adam, $\beta_F$ offers a tradeoff between outlier robustness and estimation bias.

When using optimisers like RMSProp \cite{graves2014generating} and Adam \cite{kingma2014adam}, lower variance gradients naturally accelerate convergence since these optimisers divide their step size by the standard deviation of the gradients (\cref{eq:adam_step}). Additionally, Momentum \cite{sutskever2013importance} helps these optimisers handle tricky non-linear, multivariate curvatures such as ravines. The optimisers' effectiveness is greatly reduced if the noise in the gradients overpowers the variance arising from non-linearities in the estimated moments required for these mechanisms.

Naively feeding the meta-estimated gradients to Adam is problematic; Adam computes its moment estimates, assuming the input gradients in each iteration to be independent. Meanwhile, our meta-estimator outputs an already averaged gradient (\cref{eq:recurrent_meta}) with a strong positive correlation to previous averages. Adam's moment estimates are also redundant since we already estimate the variance of our meta-estimator $\varmeta$. Therefore, we formulate the update step in terms of our estimates:
\begin{equation}
    \D \pi_{i+1} = - \eta \frac{\est_M}{\sqrt{\varmeta} + \epsilon} \text{ .}
\end{equation}
Dividing by $\sqrt{\varmeta}$ sets the step size based on our meta-estimator. As $\varmeta$ responds to changes in the estimated gradients much more quickly than Adam's second-moment estimate with the suggested $\beta_2 = 0.999$ parameter, the stability of our method may seem uncertain. We observe that the responsivity of our method actually improves convergence, especially when combined with the decoupled estimation of $\vardiff$. Optimisation speeds up quickly when low-noise gradients are available and slows down naturally when approaching a minimum.

%%%%%%%%%%%%%%%%%%%%%%%%%%%%
\section{Experiments}
\label{sec:experiments}
%%%%%%%%%%%%%%%%%%%%%%%%%%%%

We run several experiments to confirm our method's behaviour and verify its theory. We also compare our method against Adam, as it is used in state-of-the-art inverse rendering pipelines. We implement our method in Mitsuba 3 \cite{Mitsuba3} and use Path Replay Backpropagation \cite{Vicini2021PathReplay} to sample gradients computed with the unbiased Mean Relative Squared Error loss \cite{deng2022recon, Pidhorskyi2022depth}. For texture optimisation tasks, we use gradient preconditioning as proposed by \citet{Nicolet2021Large}. We compute $\diffest$ with a simplified form of the shift mapping proposed by \citet{kettunen2015gradient}, only accounting for the  BRDF sampling. While this implementation is sufficient for our proof-of-concept demonstrations, a full implementation of shift mapping can also account for changes in geometry at an insignificant cost compared to proportional samples. Unless mentioned otherwise, we tune learning rates of each method in each experiment.

\paragraph{Variance reduction without lag}
We investigate the variance reduction our method can achieve while the scene parameters are changing. We run a fixed linear interpolation of the parameters without an optimiser to prevent any effects from the feedback of the gradients.

\begin{figure}[t]
  \centering
  \includegraphics[width=\columnwidth]{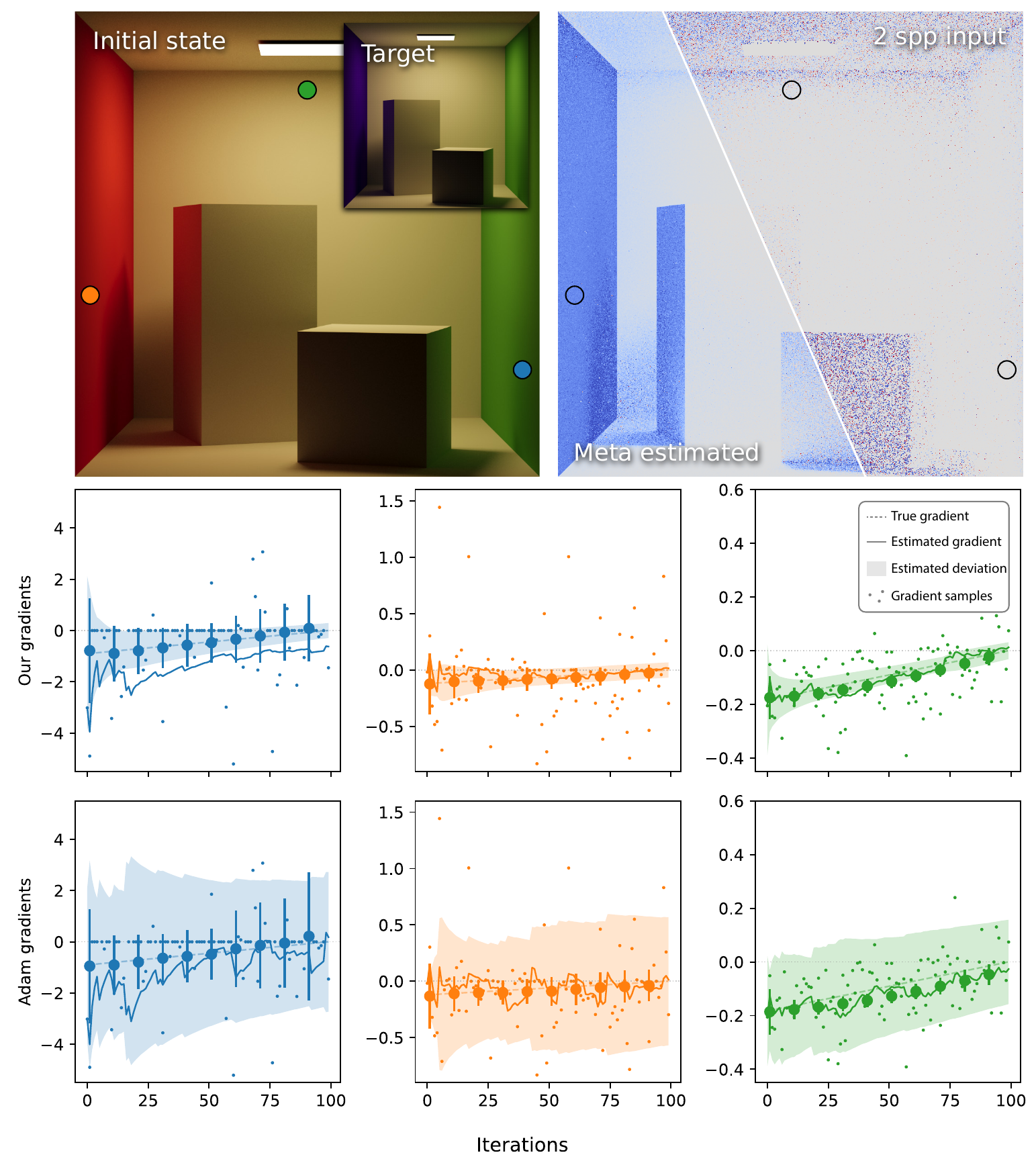}
  \caption{We estimate forward gradients of the left wall's colour's blue channel while linearly changing the scene from the initial state (top left) to the target state over 100 iterations.
  The dashed line represents the actual gradient, dots the gradient samples, the solid line the estimated gradient, and the shaded area the estimated standard deviation. \newtext{Error bars every 10 iterations show the run-to-run variation of the estimated gradient.} 
  Meta-estimation eliminates lag, improves robustness to outliers, and offers lower variance while more accurately estimating this variance. \newtext{We select the three pixels w.r.t. the actual gradient variance; blue is the noisiest, orange is at 75'th percentile, and green is the median.}}
  \label{fig:exp1}
\end{figure}

Forward gradients of several pixels in \cref{fig:exp1} show that our method avoids the lag in gradients typical of EMAs. Our meta-estimate's actual variances and estimate variances are much tighter than the estimates computed by Adam. Furthermore, our method remains more stable upon encountering outliers.

\begin{figure}[t]
  \centering
  \includegraphics[width=\columnwidth]{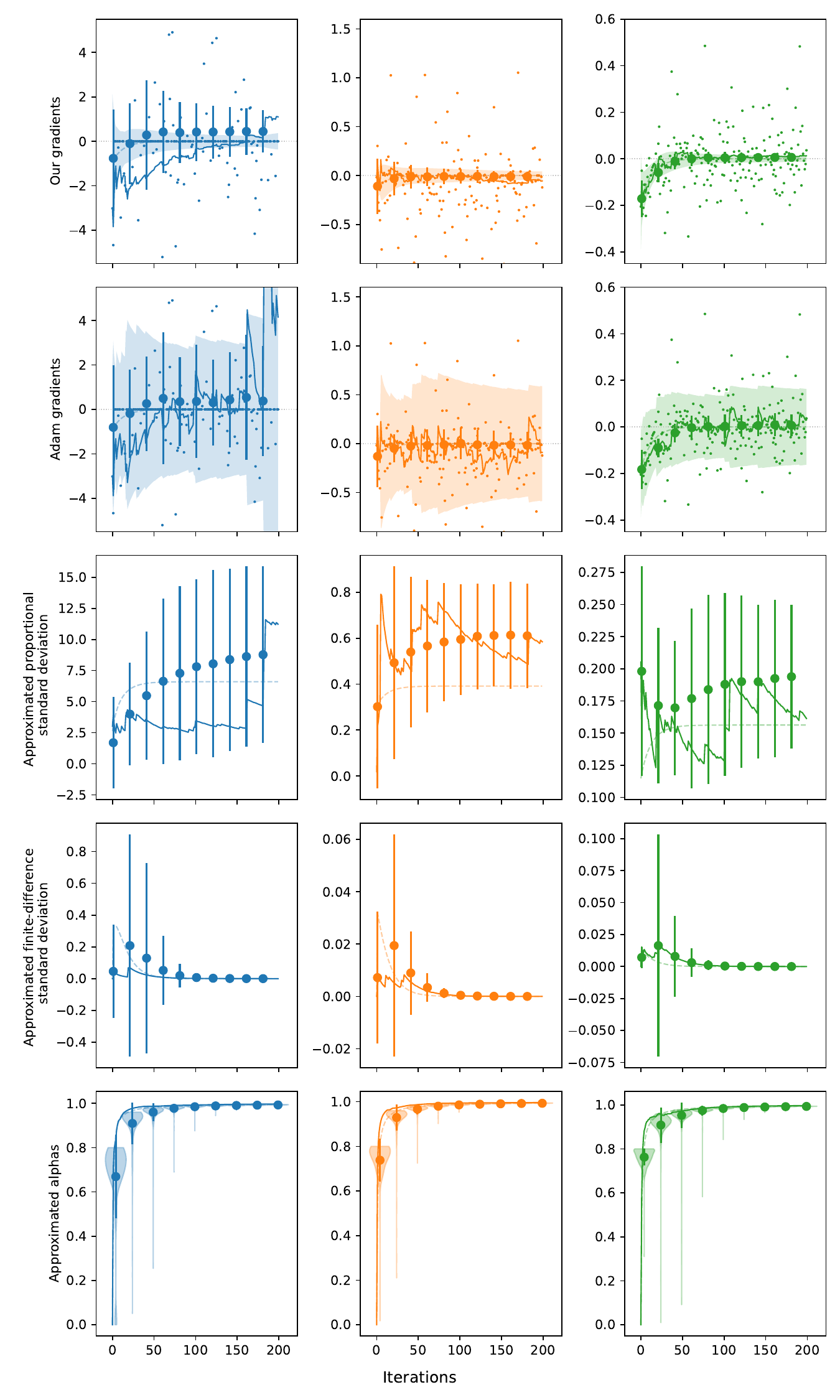}
  \caption{Following the same setup as \cref{fig:exp1}, we show the estimated gradients and standard deviations of our meta-estimator and Adam. Our meta-estimators achieve significantly lower \newtext{actual variance} in each case, while also providing much more accurate approximations. \newtext{
  Dashed lines represent actual or optimal values, solid lines a randomly sampled run, while error bars show run-to-run variation. In addition, we show violin plots for alpha, demonstrating how our method is more likely to be conservative and not to be swayed by outliers.}}
  \label{fig:components}
\end{figure}

\paragraph{Approximation accuracy} 
We repeat the previous setup in~\cref{fig:components}, only now we test an exponentially decaying change in the gradients. 
Again, our meta-estimator stays within 0.5 to 2 times its predicted standard deviation. 
As the gradients settle, our method provides a consistent variance reduction (Row 1), averaging a large number of samples wherever possible. 
Meanwhile, Adam struggles with high-variance gradients (Row 2) and is thrown off by outliers. 

\newtext{
We also show the approximated variances compared to ground truth variances computed over 1000 independent runs. 
Our approximation methods perform reasonably, only overestimating $\varest$ (Row 3). 
This overestimation results in generally conservative $\alpha_i$ values, erring on the side of robustness rather than maximising variance reduction (Row 5).
On the other hand, we approximate $\vardiff$ (Row 4) with little bias, although with often a large run-to-run variance.
}

\paragraph{Multivariate optimisation}
We simultaneously optimise an object's colour, metalness, and roughness, as shown in \cref{fig:exp2}. Thanks to our meta-estimator, our method can traverse \newtext{the loss surface} without losing past samples. Furthermore, our finite-difference estimates let our meta-estimator adjust rapidly, avoiding the overshoots typical of Momentum-based methods. Even when tuning Adam's hyperparameters for the specific problem, it can only match our method at an over 20 times increase in computational cost, not counting the time spent on hyperparameter tuning.

\paragraph{Texture optimisation}
We show a difficult texture \newtext{optimisation} case in \cref{fig:images}. Texture optimisation requires disentangling global illumination with very few gradient samples per texel. Adam can only take a few steps within a fixed budget at a high sample count, requiring a high learning rate that skips over the intricate loss surface necessary to navigate for disentangling various effects. At a lower sample count, however, Adam struggles to progress as steps devolve into a random walk as the scale of the gradients shrinks close to minima.

\newtext{
\paragraph{High-dimensional optimisation}
In~\cref{fig:astronaut}, we optimise an emissive-absorptive volume of size 256$\times$256$\times$256 voxels, totalling 70 million parameters. Perfectly fitting such a non-physical volume to rendered images is impossible. Thus, the optimiser needs to balance per-pixel losses for a good approximation, further needing to disentangle the small subset of parameters visible through any given pixel. Previous works avoid convergence to local minima by upsampling the optimised volume in several stages; our method does not need this workaround. On the other hand, \cref{fig:janga} shows that our method provides less benefit when our finite-difference estimator's sampling is too sparse across the volume.
}

\begin{figure}[t]
  \centering
  \includegraphics[width=\columnwidth]{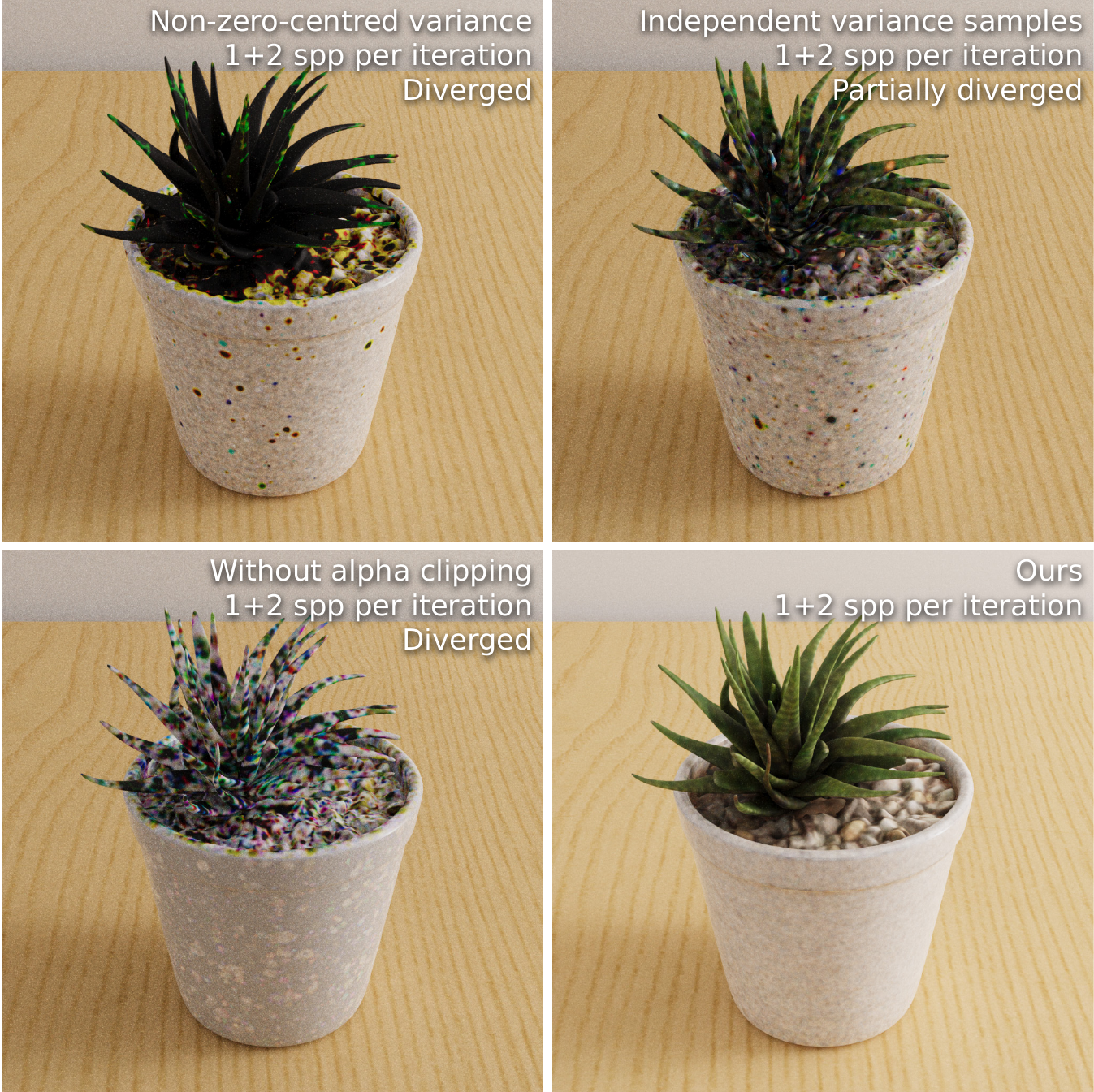}
  \caption{\newtext{
    We ablate our design choices, demonstrating their importance for robust gradient estimation and optimisation. Each modification of our method is less robust than our results in \cref{fig:images}. \cnote{Potted Plant}{James Ray Cock}
  }}
  \label{fig:ablation}
\end{figure}

\newtext{
\paragraph{Zero-centred EMAs} 
We chose to use zero-centred moving averages so that we do not need to approximate the mean of our estimators directly. This approach is generally more robust and memory efficient, though it overestimates variance at large signal-to-noise ratios. However, gradients generally have a low signal-to-noise ratio, so this tradeoff works in our favour. \Cref{fig:ablation} (top-left) demonstrates how non-zero-centred variance approximation is unstable, providing unreliable alphas, thus causing optimisation to diverge.
}

\newtext{
\paragraph{Alpha clipping} Alpha clipping helps resolve cases when our approximated variances are inaccurate, improving robustness. It may hinder the variance reduction of our method in the special case when $\varest$ is sharply decreasing over iterations. However, we have not encountered this behaviour with our tested proportional estimators. \cref{fig:ablation} (bottom-left) shows an ablation without alpha clipping for a scene from~\cref{fig:images}, demonstrating rapid divergence as the initial variance approximations are unreliable.
}

\newtext{
\paragraph{Sample reuse} We use the same samples for rendering and variance approximation. This correlation introduces some bias at the start of the optimisation process, which diminishes over time. \Cref{fig:ablation} (top-right) shows an unbiased ablation using uncorrelated samples. Although this independently approximated variance eliminates bias, it misses outliers in the samples used for gradient estimation, causing the parameters receiving these outliers to diverge.
}

\section{Limitations}
\label{sec:limitations}

Estimators $\diffest$ are not generally available for many problems. \citet{kettunen2015gradient} propose shift mapping for path tracing, which we use in our work. 
Our meta-estimators rely heavily on $\diffest$; as we recurrently sum $\vardiff$ in \cref{eq:varmeta}, it inherently bounds the variance of our meta-estimator. 
Doing so is fine as long as $\vardiff$ is quadratic w.r.t. the step size (\cref{eq:vardiff}). 
Thus, we need to ensure this property when building finite-\newtext{difference} estimators while also aiming for the lowest variance to achieve the best stability and convergence with meta-estimation.

\citet{Zeltner2021MonteCarlo} show that gradient estimators benefit from specialised differential sampling strategies. The same is true of finite-\newtext{difference} estimators; our naive toy formulation in \cref{eq:est_triv} glosses over this problem where $p(\rv{x}, \pi_i)$ is usually only optimised by importance sampling $f(\rv{x}, \pi_i)$, not the difference between $f(\rv{x}, \pi_i)$ and $f(\rv{x}, \pi_{i-1})$.

Suboptimal sampling strategies of $\est$ compound the issue. As our work focuses on gradient estimation, meaning $F_i$ are gradients, sampling of $\est$ is not yet well established. For example, \citet{Zeltner2021MonteCarlo} show the poor performance of roughness gradient estimators. \newtext{We experience these issues first-hand, as we show in \cref{fig:chalice}.} 

\section{Conclusion}
Our proposed meta-estimation technique and corresponding adaptation of the Adam update rule \newtext{can} substantially improve convergence when descending on noisy gradients, reducing computation costs by several orders of magnitude. We solve cases where low-sample-count gradients are too noisy for fast convergence while high-sample-count gradients are prohibitively expensive to compute for the required number of iterations on difficult non-linear, multivariate problems.

\paragraph{Future work} We look forward to applications of meta-estimation to various inverse Monte Carlo problems, especially as MC gradient estimators become prominent in machine learning \cite{mohamed2020monte}. Building good gradient and finite-difference estimators may seem challenging \newtext{--- and are the main limitation of our method ---} but it is undoubtedly a fruitful direction for future work. We did not investigate training deep neural networks in this work but see it as the next step once low-variance finite-difference estimators become available.

\begin{acks}
This work is supported by an academic gift from Meta. We thank the anonymous reviewers for their valuable feedback.
\end{acks}

\balance
\bibliographystyle{ACM-Reference-Format}
\bibliography{bib.bib}

\appendix
\clearpage

\begin{figure*}[t]
  \centering
  \includegraphics[width=0.67\linewidth]{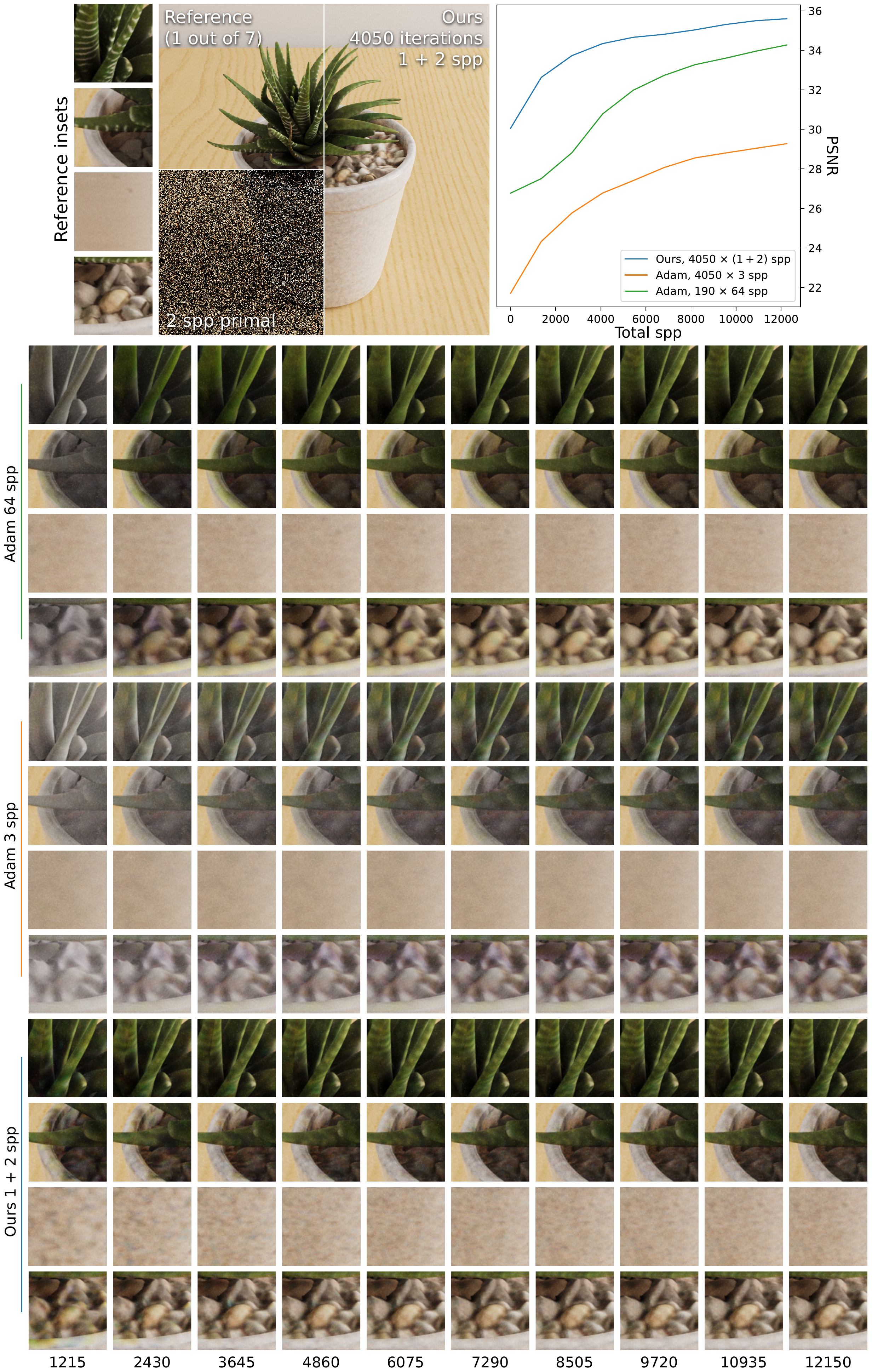}
  \caption[]{We optimise diffuse textures, placing a potted plant in the extremely noisy Veach Ajar scene. At 64 spp, Adam can only afford to take a few steps at a large learning rate and cannot match small details, further failing on complex interactions such as the leaf's reflection on the pot's side, where navigating a difficult loss surface is necessary. At 3 spp, Adam reduces to a random walk after reaching vaguely accurate parameters where gradients vanish against the high noise level. Our estimator shows good convergence. Textures remain a little blurry as a known consequence of regularisation \cite{Nicolet2021Large}. The texture of the glossy pot (third inset) shows relatively worse convergence due to the concerns discussed in \cref{sec:limitations}. \cnote{Potted Plant}{James Ray Cock}, \cnote{Veach Ajar}{Benedikt Bitterli}}
   \label{fig:images}
\end{figure*}

\begin{figure*}[t]
  \centering
  \includegraphics[width=\linewidth]{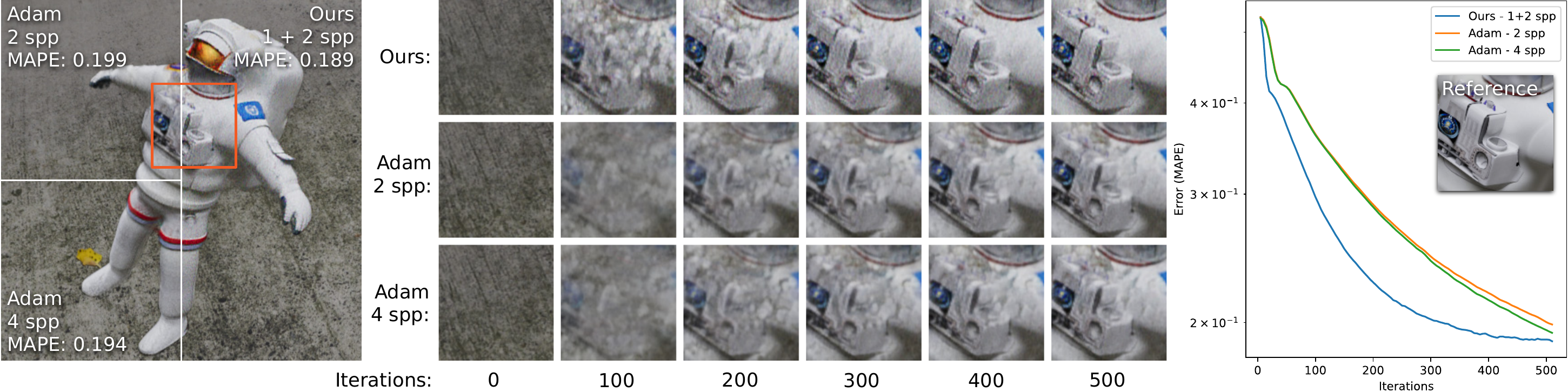}
  \caption[]{\newtext{
  We optimise emission and absorption volumes in a NeRF-like experiment. Regarding computational cost, our method lies between 2 spp and 4 spp Adam at one finite-difference and two proportional samples. However, we greatly outperform both regarding convergence and final quality. Although the final quantitative difference is smaller, we resolve jarring artefacts such as holes in the volume and overall blur. \cnote{Astronaut}{jgilhutton}
  }}
   \label{fig:astronaut}
\end{figure*}

\begin{figure*}[t]
  \centering
  \includegraphics[width=\linewidth]{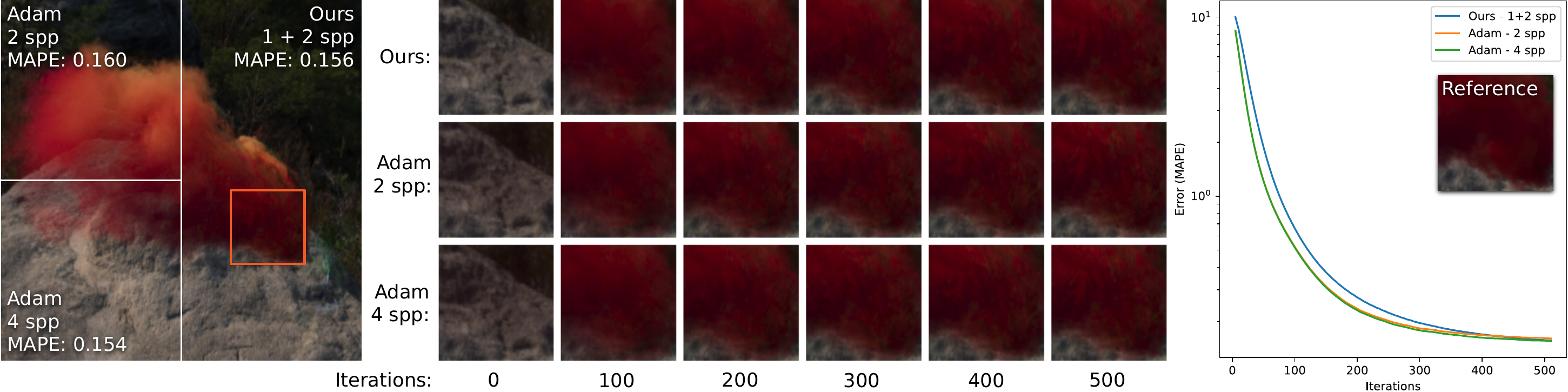}
  \caption[]{\newtext{
  Same setup as \cref{fig:astronaut}. Our method performs worse here as the volume is sparser, concentrating our finite-difference samples less. Thus, parameters at individual voxels are samples relatively sparsely, making finite-difference estimation for the whole volume difficult. \cnote{Smoke}{JangaFX Software}
  }}
   \label{fig:janga}
\end{figure*}

\begin{figure*}[t]
  \centering
  \includegraphics[width=\linewidth]{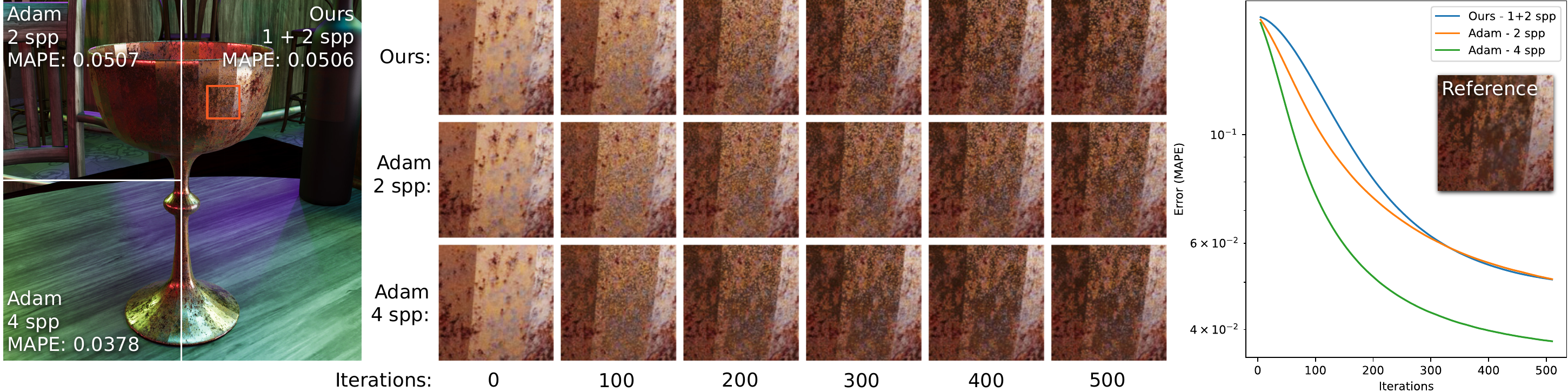}
  \caption[]{\newtext{
  We optimise the 2K roughness texture of the Chalice scene from \citet{chang2023parameter}. Using only basic shift mapping, our finite-difference estimator struggles to cope with the changes in roughness over sharp lighting. As the roughness parameter influences our sampling strategy, this is a challenging case for finite-difference estimation. \cnote{Chalice}{SusanKing}
  }}
   \label{fig:chalice}
\end{figure*}

\end{document}

% Title portion
\title{Joint Sampling and Optimisation for Inverse Rendering: Supplementary Material}

% DO NOT ENTER AUTHOR INFORMATION FOR ANONYMOUS TECHNICAL PAPER SUBMISSIONS TO SIGGRAPH 2019!
\author{Martin Balint}
\orcid{0000-0001-6689-4770}
\affiliation{%
 \institution{Max Planck Institute for Informatics}
 \streetaddress{Campus E1 4}
 \city{Saarbruecken}
 \postcode{66123}
 \country{Germany}}
\email{mbalint@mpi-inf.mpg.de}

\author{Karol Myszkowski}
\orcid{0000-0002-8505-4141}
\affiliation{%
 \institution{Max Planck Institute for Informatics}
 \streetaddress{Campus E1 4}
 \city{Saarbruecken}
 \postcode{66123}
 \country{Germany}}
\email{karol@mpi-inf.mpg.de}

\author{Hans-Peter Seidel}
\orcid{0000-0002-1343-8613}
\affiliation{%
 \institution{Max Planck Institute for Informatics}
 \streetaddress{Campus E1 4}
 \city{Saarbruecken}
 \postcode{66123}
 \country{Germany}}
\email{hpseidel@mpi-inf.mpg.de}

\author{Gurprit Singh}
\orcid{0000-0003-0970-5835}
\affiliation{%
 \institution{Max Planck Institute for Informatics}
 \streetaddress{Campus E1 4}
 \city{Saarbruecken}
 \postcode{66123}
 \country{Germany}}
\email{gsingh@mpi-inf.mpg.de}

\begin{CCSXML}
<ccs2012>
<concept>
<concept_id>10010147.10010371.10010372.10010374</concept_id>
<concept_desc>Computing methodologies~Ray tracing</concept_desc>
<concept_significance>500</concept_significance>
</concept>
</ccs2012>
\end{CCSXML}

\ccsdesc[500]{Computing methodologies~Ray tracing}
\keywords{differentiable rendering, inverse rendering, gradient estimation, gradient descent}

\maketitle

\appendix
\section{Start-up bias}
To show our start-up bias, we rewrite Equation 7 as:
\begin{multline}
    \est_M = \sum_{j=1}^{i-1} \prod_{k=j+1}^{i} [\alpha_k] (1 - \alpha_j) \< F_j \> \\
    + \prod_{k=1}^{i} [\alpha_k] \< F_0 \> + (1 - \alpha_i) \est 
    + \sum_{j=1}^{i} \prod_{k=j}^{i} [\alpha_k] \< \D F_j \>
    \text{ .}
\end{multline}
$\< F_0 \>$'s contribution diminishes over time. As $\beta_F \rightarrow 1$, the correlation between $\alpha$ and $\< F_i \>$ also diminishes, leaving us with correlation between $\alpha_k$ and $\< \D F_j \>$ as the main source of bias. As optimisation converges we expect our step size to shrink and $\vardiff$ to decline. Thus, as $\varest$ dominates Equation 8, $\alpha_k$ and $\var [\< \D F_j \>]$, become increasingly less correlated, thus, by extension, reducing correlation with $\< \D F_j \>$ (Equation 14).

Growing $\beta_F$ over the optimisation process is a simple way to make our meta-estimators consistent. However, in practice, we have found no benefit to this approach compared to setting $\beta_F$ to a fixed value that we experimentally fit to the noise characteristics of the gradient estimator $\est$.

\section{Pseudocode}
Variance approximation with zero-centred, second moment EMAs:
\begin{mintedbox}{python}
class Moment2:
    def __init__(self, shape, decay = 0.9):
        self.decay = decay
        self.count = 0
        self.m2 = 0
    
    def step(self, x):
        self.count += 1
        w = 1.0 - self.decay
        w_sum = 1.0 - self.decay ** self.count

        self.m2 = self.m2 + (w / w_sum) * (x ** 2 - self.m2)
\end{mintedbox}

Meta-estimation:
\begin{mintedbox}[escapeinside=||,mathescape=true]{python}
class Meta:
    def __init__(self, |$\eta$| = 0.001):
        self.|$\eta$| = |$\eta$|
        self.var = 0
        self.mean = 0
        self.|$\alpha_{i-1}$| = -|$\infty$|
    
    def step(self, |$\est$|, |$\diffest$|, |$\varest$|, |$\vardiff$|):
        self.mean += |$\diffest$|
        self.var += |$\vardiff$|

        |$\alpha$| = |$\varest$| / (|$\varest$| + self.var + 1e-30) # Inverse variance weighting
        |$\alpha$| = min(|$\alpha$|, 1 / (2 - self.|$\alpha_{i-1}$|)) # Alpha clipping
        self.|$\alpha_{i-1}$| = |$\alpha$|

        self.mean = |$\alpha$| * self.mean + (1 - |$\alpha$|) * |$\est$|
        self.var = |$\alpha$| ** 2 * self.var + (1 - |$\alpha$|) ** 2 * |$\varest$|

        return - self.|$\eta$| * self.mean / (sqrt(self.var) + 1e-8)
\end{mintedbox}

Optimisation loop snippet:
\begin{mintedbox}[escapeinside=||,mathescape=true]{python}
|$\varest$| = Moment2()
|$\vardiff_D$| = Moment2()
meta = Meta()

...

|$\varest$|.step(|$\est$|)
|$\vardiff_D$|.step(|$\diffest$| / sqrt(|$\D \pi_i^2$|))
|$\D \pi_{i+1}$| = meta.step(diff, |$\est$|, |$\diffest$|, |$\varest$|.m2, |$\vardiff_D$|.m2 * |$\D \pi_i^2$|)
|$\pi_{i+1}$| = |$\pi_{i}$| + |$\D \pi_{i+1}$|
\end{mintedbox}